\documentclass[summary]{gass2017}

\usepackage{algorithm}
\usepackage{algorithmic}

\title{Coalition Formation Games Based Sub-Channel Allocation for Device-to-Device Underlay mmWave Small Cells}

\author{Yong~Niu*\affref{ref1}, Han Shi\affref{ref2}, Yong~Li\affref{ref2}, Ruisi He\affref{ref1},
Zhangdui Zhong\affref{ref1}}

\affiliation{%
  \aff{ref1}{State Key Laboratory
of Rail Traffic Control and Safety, Beijing Jiaotong University, Beijing
100044, China}
  \aff{ref2}{Department
of Electronic Engineering, Tsinghua University, Beijing, China}
}

\begin{document}

\maketitle

\begin{abstract}
Small cells in the millimeter wave band densely deployed underlying the macrocell have been regarded as one of
promising candidates for the next generation mobile networks. In the user intensive region, device-to-device (D2D) communication
in physical proximity can save power and improve spectral efficiency.
In this paper, we focus on the optimal sub-channel allocation for access and D2D links in the scenario of densely deployed multiple mmWave small cells. The problem is modeled as a coalitional game to maximize the system sum rate of access and D2D links in the system.
Then we propose a coalition formation game based algorithm for sub-channel allocation. Performance evaluation results demonstrate superior performance in terms of the system sum rate compared with other practical schemes. 
\end{abstract}

\section{Introduction}

In 5G era, higher network capacity should be provided to address the challenge from explosive mobile traffic growth.
One effective way of improving network capacity is to utilize the higher frequency resources. Millimeter wave (mmWave) bands,
which has several gigahertz bandwidth, have been proposed to make a big impact in the 5G era. With huge bandwidth available,
the link transmission rate can be increased to several Gbps. Thus, bandwidth intensive applications like high-definition TV,
Augmented Reality, or virtual reality can be supported in the mmWave band. 
Due to high carrier frequency, the propagation loss is high for communications in the mmWave band.
Consequently, directional antennas are synthesized at both the transmitter and receiver to achieve high antenna gain.
Due to small wavelength, directional antennas in the form of antenna arrays can be synthesized in a small platform.
Then the transmitter and the receiver point their beams towards each other by beam training \cite{beam_training}.

Small cells are usually densely deployed in the user intensive region to serve more users with high quality services. In this situation, users are probably located in physical proximity.
Consequently, D2D communications have significant advantages to support many content-based applications \cite{Hanbook}. Due to directional communication, D2D communications in the mmWave band have less interference to access users compared with the conventional D2D communications in lower frequency bands.
In the system, there are multiple sub-channels in the mmWave band. In each small cell, multiple access users in different sub-channels can be supported simultaneously by the base station.
Therefore, how to allocate sub-channels to access links and D2D links in this scenario to reduce interference and maximize network capacity becomes
a key problem.

In this paper, we study the problem of optimal sub-channel allocation for mmWave small cells densely deployed. We address this problem using game theory, and the coalition formation games
are utilized to model the sub-channel allocation problem. In a coalition game, members form coalitions to improve system performance. In our problem, links occupying the same sub-channel are formed as a coalition, and the sum rate of links in the coalition is the total utility of this coalition. The sum rate of all links in the system is the total utility all coalitions try to maximize.

\section{System Model}

We consider the scenario of small cells in the mmWave bands densely deployed. In each small cell, mobile
users are associated with corresponding base stations, and the access links are in the mmWave bands. Besides, we also enable D2D communications in the mmWave bands within each small cell and between small cells.
There are multiple sub-channels in the mmWave bands, and the access links between users and base stations, and the D2D links between users are all in the mmWave bands. 
Multiple access links in different mmWave sub-channels can be supported simultaneously in the same small cell.

The link from nodes $i$ to $j$ is denoted by $(i,j)$. In the mmWave bands, we assume node $i$ and node $j$
point their directional beams towards each other for directional transmissions. The transmit antenna
gain of node $i$ in the direction of $i \to j$ is denoted by ${G_t}(i,j)$, and the receive antenna gain of node $j$ in the direction
of $i \to j$ by ${G_r}(i,j)$. If we denote the distance between nodes $i$ and $j$ by $l_{ij}$, then the received power
at node $j$ from node $i$ can be obtained according to the path loss model \cite{Qiao} as
\begin{equation}
{P_r}(i,j) = {k_0}{G_t}(i,j){G_r}(i,j)l_{ij}^{ - n}{P_t},
\end{equation}
where $P_t$ is the transmission power, $n$ is the path loss exponent, and ${k_0}$ is a
constant proportional to ${(\frac{\lambda }{{4\pi }})^2}$ ($\lambda $ is the wavelength).
For two links $(i,j)$ and $(u,v)$ in the same sub-channel, the interference power at node $j$ from node $u$ can be obtained as
\begin{equation}
P_{uvij} =\rho k_0{G_t}(u,j){G_r}(u,j){l_{uj}}^{ - n }{P_t},
\end{equation}
where $\rho$ is the multi-user
interference (MUI) factor and relates to the cross correlation of signals from different links \cite{Qiao}.
Then we can obtain the received signal to interference plus noise ratio (SINR) at receiver $j$ as
\begin{equation}
{\Gamma_{ij}} = \frac{{{k_0}{G_t}(i,j){G_r}(i,j){l_{ij}}^{ - n }P_t}}{{{N_0}{W_0} + \rho \sum\limits_{(u,v)} {k_0{G_t}(u,j){G_r}(u,j){l_{uj}}^{ - n }{P_t}} }},
\end{equation}
where ${{N_0}}$ denotes the one-sided
power spectra density of white Gaussian noise.
Due to lack of multipath effect for directional mmWave links,
the achievable transmission rate of link $(i,j)$ can be obtained according to the Shannon's
channel capacity as
\begin{equation}
\begin{array}{l}\begin{aligned}
&{R_{ij}} = \eta {W_0}\cdot\\&{\log _2}\left(1 + \frac{{{k_0}{G_t}(i,j){G_r}(i,j){l_{ij}}^{ - n }P_t}}{{{N_0}{W_0} + \rho \sum\limits_{(u,v)} {k_0{G_t}(u,j){G_r}(u,j){l_{uj}}^{ - n }{P_t}} }}\right),
\end{aligned}\end{array}\label{rate}
\end{equation}
where $\eta \in (0,1)$ is the efficiency of the transceiver design \cite{Qiao}.

\section{Coalition Formation Game}

In this section, we model the sub-channel allocation problem as a coalitional game, where links as the game players tend to
form coalitions to improve the system utility in terms of the system sum rate.

\subsection{Coalitional Game}

We use $a$ to denote one access link, and $t_a$ and $r_a$ are the transmitter and receiver of link $a$, respectively.
We use $d$ to denote one D2D link, and $t_d$ and $r_d$ are the transmitter and receiver of link $d$, respectively.
We consider the uplink access links, and D2D links share sub-channels with the uplink access links.
One access link's sub-channel can be shared with multiple D2D links to maximize spectral efficiency, and one D2D link occupies at most
one sub-channel. We also assume one access link occupies at most one sub-channel. Under the same base station, multiple access links in different sub-channels can be supported. We denote the set of sub-channels by $C$, and denote one sub-channel by $c\in C$.
We denote the set of access links by $A$, and the set of D2D links by $D$.

In our presented problem above, there are $|A|$ access links and $|D|$ D2D links, and they share $|C|$ sub-channels to achieve higher system performance in terms of the system sum rate. In the following, we give the definition of a coalitional game.
A coalitional game with the transferable utility is defined by a pair $(\Omega ,R)$, where $\Omega$ is the set of game players, and
$R$ is a function over the real line such that for every coalition $S \subseteq \Omega $, $R(S)$ is a real number describing the amount of value
that coalition $S$ receives that can be distributed in any arbitrary manner among the members of $S$.

We can observe that with more links occupying the same sub-channel, there will be more interference between links, and the system sum rate will decrease. Besides, access links under the same base station cannot occupy the same sub-channel. Therefore, there is no motivation to form as a grand
coalition for occupying only one sub-channel. In fact, links will form as independent as possible disjoint coalitions in different sub-channels to maximize the system sum rate.
Considering $|C|$ sub-channels, the links can form $|C|$ coalitions with links occupying the same sub-channel as a coalition.
We denote the coalitions as $\Omega={S_1} \cup {S_2} \cup  \cdots  \cup {S_{|C|}}$, where ${S_x} \cap {S_{x'}} = \emptyset $ for any $x \ne x'$.
With links in $S_c$ sharing the sub-channel $c\in C$, we can obtain the transmission rate of links $a\in S_c$ as

\begin{equation}
{R_{a}} = \eta {W_0}{\log _2}\left(1 +\frac{{{k_0}{G_t}(t_a,r_a){G_r}(t_a,r_a){l_{t_ar_a}}^{ - n }P_t}}{{{N_0}{W_0} + \sum\limits_{a' \in S_c\backslash a} {{I_{a',a}} + \sum\limits_{d' \in S_c} {{I_{d',a}}}}}}\right),
\end{equation}
$I_{a',a}$ is the interference power from link $a'$ to $a$. $I_{d',a}$, $I_{a',d}$, and $I_{d',d}$ are similar.
The transmission rate of link $d\in S_c$ can be obtained accordingly.
Thus, the sum rate of links in $S_c$ can be obtained as
\begin{equation}
R(S_c)=\sum\limits_{a \in {S_c}} {{R_a}}  + \sum\limits_{d \in {S_c}} {{R_d}} .
\end{equation}
Therefore, the sub-channel allocation problem can be modeled as a coalitional game with the transferable utility, where $\Omega$ is the set of access and D2D links. These links tend to form coalitions in different sub-channels to maximize the utility of all coalitions. 

\textbf{Coalitional Game for Sub-channel Allocation}:
The coalitional game with transferable utility for sub-channel allocation of uplink access links and D2D links is defined by a pair $(\Omega,R)$,
and the game formation is as follows.
\begin{itemize}
  \item \emph{Players}: the set of access and D2D links $\Omega=A \cup D$.
  \item \emph{Transferable Utility}: $R(S_c)$ is the value for each coalition $S_c \subseteq \Omega$, which is a transferable utility for members in $S_c$.
  \item \emph{Coalition Partition}: The set of players $\Omega$ is partitioned into $|C|$ coalitions, i.e., $\Omega={S_1} \cup {S_2} \cup  \cdots  \cup {S_{|C|}}$. ${S_x} \cap {S_{x'}} = \emptyset $ for any $x \ne x'$.
  \item \emph{Strategy}: The players make a decision to join or leave a coalition based on the utilities of the original coalition and the new coalition.
\end{itemize}

\subsection{Coalition Formation Algorithm}

To maximize the system sum rate, preference relation should be well defined for players to decide whether to
join or leave a coalition.
Since we try to maximize the sum rate of links in $\Omega$, we adopt the utilitarian order
in \cite{CG-tutorial}, i.e., a group of players prefers to organize themselves into a collection of coalitions ${\bf{R}} = \{ {R_1}, \ldots ,{R_k}\}$
instead of ${\bf{S}} = \{ {S_1}, \ldots ,{S_l}\} $, if the total utility achieved by ${\bf{R}}$ is strictly greater than by ${\bf{S}}$, i.e.,
$\sum\limits_{i = 1}^k {v({R_i}) > \sum\limits_{i = 1}^l {v({S_i})} } $, which is very suitable for coalitional games with transferable utility.
For a partition $\Pi= \{ {S_1}, \ldots ,{S_t}\} $ ($1 \le t \le |C|$) of the player set $\Omega$, its total utility can be expressed as
$R(\Pi ) = \sum\limits_{i = 1}^t {R({S_i})} $.
Therefore, partition $\Pi$ is preferred over $\Pi'$ for maximizing the total utility if $R(\Pi )>R(\Pi' )$.

In the following, we define the preference relation for each player $l\in \Omega$.
For any player $l\in \Omega$, a preference relation ${ \succ _l}$ is defined as a complete, reflexive, and transitive
binary relation over the set of all coalitions that player $l$ may form, i.e., $\{ {S_c} \subseteq \Omega :l \in {S_c}\} $.
For any player $l\in \Omega$, ${S_1}{ \succ _l}{S_2}$ means player $l$ prefers being a member of coalition $S_1$ over
being a member of coalition $S_2$.
Thus, the preference relation ${ \succ _l}$ with $l\in{S_1}$ and $l\in{S_2}$ is quantified as follows:
\begin{equation}
{S_1}{ \succ _l}{S_2} \Leftrightarrow R({S_1}) + R({S_2}\backslash \{ l\} ) > R({S_2}) + R({S_1}\backslash \{ l\} ).\label{preference}
\end{equation}
This definition implies that player $l$ prefers being a member of $S_1$ over $S_2$ only when there is an increase in the total utility of members in $S_1$ and $S_2$. Similarly, we define the preference relation ${\succeq _l}$ as follows.
\begin{equation}
{S_1}{ \succeq _l}{S_2} \Leftrightarrow R({S_1}) + R({S_2}\backslash \{ l\} )  \ge  R({S_2}) + R({S_1}\backslash \{ l\} ).\label{preference2}
\end{equation}

In the following, we give the definition of the set of base stations of access links in each coalition, and based on this definition, we define the switch operation in our coalition game.
Given a coalition $S_c$, we define $B_c$ as the set of base stations of access links in $S_c$, i.e., ${B_c} = \{ {r_a}| a \in {S_c}\} $.
Since the access links under the same base station cannot occupy the same sub-channel, the base station of access links joining $S_c$
should be different from those in $B_c$. In other words, if access link $a$ want to join $S_c$, then $r_a \notin B_c$ should hold.

\textbf{Switch Operation}:
Given a partition of $\Pi= \{ {S_1}, \ldots ,{S_t}\} $ ($1 \le t \le |C|$) of the player set $\Omega$, if link $l\in \Omega$ performs a switch operation from $S_m$
to $S_k\in \Pi \cup \{ \emptyset \} $, $S_k \ne S_m$, $r_l \notin B_k \;{\rm{if}} \;l\in A$, then the current partition $\Pi$ of $\Omega$ is modified into a new partition $\Pi '$ such that
$\Pi ' = (\Pi \backslash \{ {S_m},{S_k}\} ) \cup \{ {S_m}\backslash \{ l\} \}  \cup \{ {S_k} \cup \{ l\} \} $.

Then we can obtain the basic rules for switch operations to maximize system sum rate.
\textbf{Switching Rules}: Given a partition of $\Pi= \{ {S_1}, \ldots ,{S_t}\} $ ($1 \le t \le |C|$) of the player set $\Omega$,
a switch operation from $S_m$ to $S_k\in \Pi \cup \{ \emptyset \} $, $S_k \ne S_m$, $r_l \notin B_k \;{\rm{if}} \;l\in A$ is allowed for any player $l\in \Omega$, if and only if
${S_k} \cup \{ l\} { \succ _l}{{\rm{S}}_m}$.
Each link $l\in \Omega$ can leave its current coalition $S_m$ to join another coalition $S_k\in \Pi \cup \{ \emptyset \} $ if the new coalition
${S_k} \cup \{ l\}$ is strictly preferred over $S_m$ through the preference relation defined in (\ref{preference}).

We summarize the coalition formation game for sub-channel allocation in Algorithm \ref{alg:CG}. As shown in lines 9--16, when the first switch operation fails, we further examine the second switch operation, and if the
partition after the second switch operation has higher total utility than the current partition, these two switch operations will be performed, and the current partition will be updated as the partition after the second switch operation.

\begin{algorithm}[t]
\caption{Coalition Formation Algorithm for Sub-channel Allocation}
\label{alg:CG}
\begin{algorithmic}[1]
\STATE {Initialize the system by any random partition ${\Pi _{ini}}$;}
\STATE {Set the current partition ${\Pi _{cur}}={\Pi _{ini}}$;}
\REPEAT
\STATE Uniformly randomly choose one link $l\in \Omega$, and denote its current coalition as $S_m \in {\Pi _{cur}}$;
\STATE Uniformly randomly choose another coalition $S_k\in {\Pi _{cur}} \cup \{ \emptyset \} $, $S_k \ne S_m$, $r_l \notin B_k \;{\rm{if}} \;l\in A$;
\IF{the switch operation from $S_m$ to $S_k$ satisfying ${S_k} \cup \{ l\} { \succ _l}{{\rm{S}}_m}$}
\STATE Link $l$ leaves its current coalition $S_m$, and joins the new coalition $S_k$;
\STATE Update the current partition ${\Pi _{cur}}$   as \\$({\Pi _{cur}}\backslash \{ {S_m},{S_k}\} ) \cup \{ {S_m}\backslash \{ l\} \}  \cup \{ {S_k} \cup \{ l\} \}  \to {\Pi _{cur}}$;
\ELSE
\STATE Obtain the temporary partition ${\Pi _{tmp}}$ as \\$({\Pi _{cur}}\backslash \{ {S_m},{S_k}\} ) \cup \{ {S_m}\backslash \{ l\} \}  \cup \{ {S_k} \cup \{ l\} \}  \to {\Pi _{tmp}}$;
\STATE Uniformly randomly choose one link $l'\in \Omega$, and denote its current coalition as $S_{m'} \in {\Pi _{tmp}}$;
\STATE Uniformly randomly choose another coalition $S_{k'}\in {\Pi _{tmp}} \cup \{ \emptyset \} $, $S_{k'} \ne S_{m'}$, $r_{l'} \notin B_{k'} \;{\rm{if}} \;l'\in A$;
\STATE Obtain the partition ${\Pi' _{tmp}}$ as \\$({\Pi _{tmp}}\backslash \{ {S_{m'}},{S_{k'}}\} ) \cup \{ {S_{m'}}\backslash \{ l'\} \}  \cup \{ {S_{k'}} \cup \{ l'\} \}  \to {\Pi' _{tmp}}$;
\IF{$R(\Pi'_{tmp} )>R(\Pi_{cur} )$}
\STATE Update the current partition ${\Pi _{cur}}$ as $\Pi'_{tmp} \to {\Pi _{cur}}$;
\ENDIF
\ENDIF
\UNTIL{the partition converges to a final Nash-stable partition ${\Pi _{fin}}$}
\end{algorithmic}
\end{algorithm}

\section{Performance Evaluation}
\subsection{Simulation Setup}

In the system, we consider the scenario of multiple mmWave small cells densely deployed, and D2D communications between user equipments (UEs) are enabled to share the sub-channels with access users. The mmWave small cells are randomly distributed in a circular region of radius $R=100m$.
The maximum distance of D2D links is 5m, and D2D links are randomly generated.
The directional antenna model is from IEEE 802.15.3c
 with a main lobe of the Gaussian form in linear scale and constant level of side lobes \cite{chen_2}.The parameters of the simulated mmWave small cells are summarized in
 Table~\ref{tab:para-CG}.
 
\begin{table}[htbp]
\begin{center}
\caption{Simulation Parameters.}
\label{tab:para-CG} 
\begin{tabular}{ccc}
\hline
\textbf{Parameter} & \textbf{Symbol} & \textbf{Value} \\ \hline
 Sub-channel bandwidth & $W_0$ & 540 MHz \\
 Background noise &$N_0$& -134dBm/MHz \\
 Path loss exponent & $n$ & 2 \\
 MUI factor & $\rho$ & 1\\
 Transmission power & $P_t$ & $30$dBm \\
 Maximum distance of D2D & $d$ & 5m\\
\hline
\end{tabular}
\end{center}
\vspace*{-5mm}
\end{table}

To simplify the denotation, we denote our coalition formation algorithm for sub-channel allocation by \textbf{CG}.
To show the advantages of our sub-channel allocation algorithm, we compare our scheme with other three schemes.
 1)~\textbf{RA}: Random Allocation, where the sub-channels are allocated to each access or D2D link randomly.
 2)~\textbf{PCG}: Partial Coalition Game based algorithm, where the sub-channels are allocated to access links randomly, and the sub-channels are allocated to D2D links by the coalition formation algorithm.
In the performance evaluation, we investigate the system sum rate, which is the sum of transmission rates of all D2D links and access links in the system.
 
\subsection{System Sum Rate}

\begin{figure}[tb]
\vspace*{-1mm}
\begin{minipage}[t]{1\linewidth}
\centering
\includegraphics[width=0.90\columnwidth]{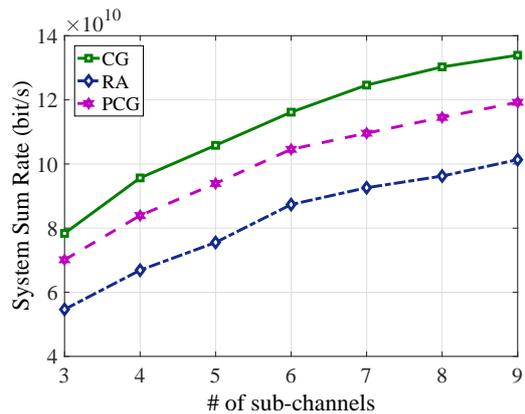}
\end{minipage}%
\vspace*{-3mm}
\caption{System sum rates of different resource allocation algorithms under different number of sub-channels.}
\label{fig:SSR}
\vspace*{-1mm}
\end{figure}

In Fig. \ref{fig:SSR}, we plot the comparison of the system sum rates of different resource allocation algorithms under different numbers of sub-channels.
There are three mmWave small cells in the system, and 15 access links and 5 D2D links are considered. From the results, we can observe that our scheme has the highest system sum rate among three schemes. When the number of sub-channels is 9, our scheme improves the system sum rate by about 32.2\% compared with the random allocation scheme, and by about 12.3\% compared with the PCG scheme.
With the increase of sub-channels, a higher system sum rate can be achieved for all schemes. With more sub-channels, there is less interference between links, and transmission rate of each link can be higher.
The gap between CG and PCG demonstrates the advantages of including access links into the coalition formation game.

\begin{figure}[tb]
\vspace*{-1mm}
\begin{minipage}[t]{1\linewidth}
\centering
\includegraphics[width=0.90\columnwidth]{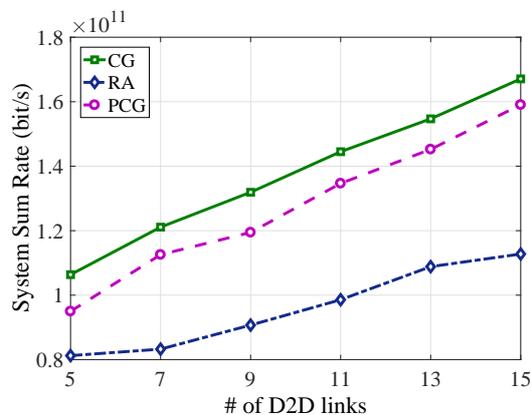}
\end{minipage}%
\vspace*{-3mm}
\caption{System sum rates of different resource allocation algorithms under different number of D2D links.}
\label{fig:SSR-D2D}
\vspace*{-1mm}
\end{figure}

In Fig. \ref{fig:SSR-D2D}, we plot the comparison of the system sum rates of different resource allocation algorithms under different numbers of D2D links. 
We can observe that our scheme also performs best in terms of the system sum rate among three schemes. With the increase of D2D links, the system sum rates of different schemes increase. With more D2D links sharing the spectrum resources of access links, a higher system sum rate can be achieved.
When the number of D2D links is 15, our scheme improve the system sum rate by about 48.2\% compared with the random allocation scheme.

%

\end{document}